# Realizing Chemical Codoping in Oxide Semiconductors

Fang Wang[1,2], Y. Y. Sun[3], John B. Hatch[2], Hui Xing[2], Hongwang Zhang[2], Xiaohong Xu[1], Hong Luo[2], S. B. Zhang[3,*], and Hao Zeng[2,*]

[1]*School of Chemistry and Materials Science, Shanxi Normal University, Linfen 041004, China*

[2]*Department of Physics, University at Buffalo, SUNY Buffalo, NY 14260, USA*

[3]*Department of Physics, Applied Physics, and Astronomy, Rensselaer Polytechnic Institute, Troy, NY 12180, USA*

*Emails: zhangs9@rpi.edu; haozeng@buffalo.edu

ABSTRACT

We demonstrate experimentally a chemical codoping approach that would simultaneously narrow the band gap and control the band edge positions of oxide semiconductors. Using $TiO_2$ as an example, we show that a sequential doping scheme with nitrogen (N) leading the way, followed by phosphorous (P), is crucial for the incorporation of both N and P into the anion sites. Various characterization techniques confirm the formation of the N-P bonds, and as a consequence of the chemical codoping, the band gap of the $TiO_2$ is reduced from 3.0 eV to 1.8 eV. The realization of chemical codoping could be an important step forward in improving the general performance of electronic and optoelectronic materials and devices.

The ability to control impurity doping in semiconductor materials lays the foundation of modern electronics. A successful example is the realization of Mg doping in GaN [1] leading to the birth of blue-LEDs [2]. Another example is controlled production of N-*V* centers in diamonds, which may enable quantum computing [3]. Novel approaches to impurity doping can enable ground-breaking applications and are highly desirable. Recently, it has been proposed theoretically that oxide semiconductors such as $TiO_2$ can be codoped with a pair of anion dopants that form strong chemical bonds [4,5]. This approach termed *chemical codoping* has several advantages: 1) The strong bonding between dopants significantly reduces the formation energy, so that higher doping concentration may be achieved. 2) The codopants can be so chosen to eliminate half-occupied states serving as recombination centers detrimental to carrier extraction. Chemical codoping, however, is counterintuitive because impurities with the same charge typically repel each other [6].

$TiO_2$ is a wide gap semiconductor commonly used for solar energy harvesting applications. However, its band gap is too large for visible light absorption. Earlier efforts for band gap engineering including monodoping of metal [7,8] and non-metal elements [9,10], passivated codoping [11], non-compensated codoping [12], and surface defect engineering [13], have made great progress to broaden its absorption range. Another essential requirement for using $TiO_2$ in photocatalytic applications is the correct position of the band edge states for carrier extraction. For spontaneous water splitting, the optical gap of $TiO_2$ should straddle the redox potentials of $H_2O$ [14,15]. Since the conduction band minimum (CBM) of pristine $TiO_2$ is only slightly higher than the $H^+/H_2$ reduction potential while its valence band maximum (VBM) is much lower than the $OH^-/O_2$ oxidation potential, the gap narrowing should be achieved by shifting the VBM upwards. The impact of developing a general approach to simultaneously engineer the band gap and band edge alignment of semiconductors is however, not limited to photocatalysis, but also for many other applications ranging from optoelectronics, spintronics, and multiferroicity [16,17] that require carrier extraction or dopant ionization.

Herein we demonstrate experimentally the realization of chemical codoping to engineer the band gap and band alignment of semiconductor oxides simultaneously, using $TiO_2$ as an example. Realizing chemical codoping is nontrivial. A single precursor containing both anion dopants would not work since the covalent bonding between them would be so strong that it prevents the incorporation of both anions. To overcome this difficulty, a sequential doping strategy was devised to enhance dopant incorporation and the ability to form new bonds. Anatase structured $TiO_2$ was codoped with nitrogen (N) and phosphorous (P) anions leading to the formation of new chemical bonds, which lowers its band gap from 3.0 eV to 1.8 eV. This value is much lower than that obtained by N (2.64 eV) [18] or P (2.91 eV) [19] monodoping, and rivals the result reported for non-compensated codoping [12].

Chemical synthesis based on hydrolysis followed by hydrothermal treatment was used to produce $TiO_2$ nanoparticles [20]. Trioctylphosphine (TOP) and urea were chosen as precursors for phosphorus and nitrogen doping [21,22]. The two elements were sequentially doped: $TiO_2$ doped with N(P) first followed by P(N) doping was labeled as NP-$TiO_2$ (PN-$TiO_2$); while samples doped with P(N) only were labeled as P-$TiO_2$ (N-$TiO_2$). Briefly, N-$TiO_2$ was obtained by dissolving titanium tetraisopropoxide in ethanol; urea dissolved in deionized water was then added dropwise to form amorphous $TiO_2$ nanoparticles. The product was then sealed in an autoclave and heated at 175 °C for 3 h to obtain the anatase structure. To obtain NP-$TiO_2$, N-$TiO_2$ was mixed with TOP and heated at the reflux temperature for 12 h. Uniform and nearly spherical $TiO_2$ particles with sizes of about 1 μm have been obtained. These particles contain nanosized crystallites ranging from 5 to 10 nm. The doping concentration and valence states of the dopants were determined by energy dispersive X-ray spectroscopy (EDX) and X-ray photoelectron spectroscopy (XPS). The optical band gaps were measured by UV-vis absorption spectra and scanning tunneling spectroscopy (STS) was used to map the spatial distribution and obtain the average band gaps. First-principles calculations were based on density functional theory (DFT) using the Heyd-Scuseria-Ernzerh of (HSE) hybrid functional [23], as implemented in the VASP program [24]. A supercell of 96 atoms (32 Ti and 64 O) was used to model anatase $TiO_2$. In the case of chemical codoping, two neighboring O atoms were replaced by N and P, respectively. The atomic structures were optimized until the calculated forces on all atoms were smaller than 0.05 eV/Å.

N and P were chosen as codopants, which according to first-principles calculation could form a strong chemical bond when both substitute for O [4]. Figure 1(a) shows the atomic structure of the N-P pair substituting for two O atoms and Figure 1(b) illustrates the electronic structure and bonding mechanism of the N-P pair. All the spin orbitals of $N_O$ and $P_O$ are involved in the bonding between the N-P pair. After the hybridization, only two fully occupied orbitals are left in the band gap, as shown in the middle part of Figure 1(b), representing the new VBM. Both gap states exhibit anti-bonding character between P *3p* and N *2p*, as can be seen in Figures 1(c) and 1(d), and are higher in energy than the occupied states in either monodoping case, resulting in more efficient band gap reduction. The bonding orbitals are pushed down into the valence band and the anti-bonding orbital is pushed up into the conduction band. The electronic structure analysis suggests that the N-P pair forms a triple bond, which is responsible for the significant drop of the formation energy for the $P_O$ defect next to $N_O$ (as discussed below).

While N monodoping has been done extensively, there are few reports of P doped $TiO_2$ [25,26]. We are not aware of work on P anion doping. The difficulty of P anion doping stems from the mismatch between the ionic radii of P (1.06 Å) and O (0.73 Å). In the few reports of (N,P) codoped $TiO_2$, the P dopants act as cations replacing Ti ions or as interstitials. With $P^{5+}$ substituting $Ti^{4+}$ and $N^{3-}$ substituting $O^{2-}$, this can be considered as passivated codoping [11], since P is a single donor and N is a single acceptor [27,28]. Our DFT calculation suggests that even though the band gap can be reduced to ~2.5 eV, the improvement on light absorption is marginal, which is verified by our experimental results. A well-designed doping strategy is therefore essential to ensure N and P incorporated as anions. Our calculation shows that the presence of N can lower the incorporation energy of P in $TiO_2$ by about 3.4 eV as a result of the formation of N-P bond. We therefore adopted a sequential doping scheme, where N was doped first, followed by P doping. Such a scheme is also favored considering the doping kinetics. It is known that $TiO_2$ nanoparticles synthesized by hydrothermal methods are *n*-type due to O vacancies. We suggest that if N is doped substitutionally at the O sites first, they are likely attracted to the vicinities of the O vacancies due to donor-acceptor attractions. This can facilitate the N-P pair formation if P dopants later assume the O vacancy sites.

The valence states and bonding of N, P, O and Ti in undoped and codoped $TiO_2$ were analyzed by XPS, as shown in Figure 2. N *1s* and P *2p* peaks emerge upon doping. These peak positions shift according to the various doping schemes. The chemical shifts are consistent with expected oxidation states and chemical bonding of the dopants, as discussed below. In Figure 2(a), an N *1s* peak centered at 399.6 eV is observed for N-$TiO_2$ with N concentration of 2.4%. When P doping is carried out after N doping to form NP-$TiO_2$ (N: 1.0%, P: 3.7%), the N *1s* peak position shifts to a slightly lower binding energy of 398.9 eV. The N *1s* binding energies in both samples are lower than that of N-O adsorption (400-406 eV) [29,30], and consistent with reported values in earlier work on N substitutional doping [31]. This clearly shows that N substitutionally dopes the O sites for both N-TiO$_2$ and NP-TiO$_2$ samples. A 0.7 eV red shift in the binding energy of NP-TiO$_2$ compared to that of N-TiO$_2$ is consistent with the formation of chemical bonding between N and P, which leads to partial electron transfer from P to N.

Evidence of P ions substitutionally doping at the O sites can be obtained by examining the P *2p* XPS peaks. Figure 2(b) presents the P *2p* XPS spectra of NP-TiO$_2$, P-TiO$_2$, and PN-TiO$_2$. A peak centered at about 133.3 eV is found for P-TiO$_2$, which indicates that P exists as P$^{5+}$ [32,33]. The P atoms were probably incorporated as cations and replaced Ti ions, or adsorbed at the surface of the particles as phosphates. For NP-TiO$_2$, on the other hand, the P *2p* peak occurs at 132.2 eV. This 1.1 eV red shift in the binding energy of P dopants in NP-TiO$_2$ strongly suggests that the valence of P dopants is lower than that of P-TiO$_2$. Considering that the peak width of 1.8 eV in NP-TiO$_2$ is 0.3 eV broader than that in P-TiO$_2$, and that the doping concentration is 1% N and 3.7% P, it is reasonable that the peak is actually a superposition of two: one from P dopants occupying the O sites, and the other from P$^{5+}$ substituting Ti$^{4+}$ ions. Due to insufficient spectral resolution, a reliable separation of the two peaks is difficult.

To provide further evidence for N-P chemical codoping, a control experiment using identical precursors and synthetic parameters was carried out, where P was doped first followed by N doping. From the XPS results, it is seen that the P *2p* peak of PN-TiO$_2$ is located at 133.3 eV, identical to that in P-TiO$_2$, suggesting that P exists as P$^{5+}$. The inability to obtain P anion doping in PN-TiO$_2$ suggests that N dopants play a critical role in P anion doping. Our DFT calculation shows that without the presence of N, the formation energy for a P-on-Ti (P$_{Ti}$) defect is lower than that for a P-on-O (P$_O$) defect over a wide range of O chemical potential. In the presence of N dopants, however, the formation energy of a $P_O$ defect in the neighborhood of a $N_O$ defect can be lowered by 3.4 eV as a result of the new chemical bonding. From these experimental and theoretical investigations, it is concluded that the only possibility for both N and P to be incorporated as anions is for them to form chemical bonds, which is the case in NP-$TiO_2$.

It is expected that the chemical codoping should lead to significant band gap reduction [4,5]. To verify this, the UV-vis spectra of $TiO_2$, N-$TiO_2$, P-$TiO_2$ and NP-$TiO_2$ samples are shown in Figure 3(a). While the optical absorption edges of the N-, P-, and NP-doped samples all shift to lower energy than that of the undoped sample, P-doped $TiO_2$ shows only minimal red shift, with an onset wavelength at about 400 nm. The UV-vis spectrum for N-$TiO_2$ shows larger red shift consistent with earlier reports [9,34]. For NP-$TiO_2$, the absorption edge is red shifted from 370 nm to 680 nm, which corresponds to a band gap of ~1.85 eV. NP-$TiO_2$ also demonstrates the strongest absorption in the visible range between 400 and 700 nm. Figure 3(b) shows the calculated imaginary part ($\varepsilon_2$) of the dielectric constant, which determines the optical absorption. It can be seen that $\varepsilon_2$ is significantly increased from that of undoped $TiO_2$ in the visible region. Meanwhile, the change of $\varepsilon_2$ for (N, P) compensated codoping with P on Ti sites is insignificant, which is consistent with the experimental result that PN-$TiO_2$ shows no improvement from pure $TiO_2$ in optical absorption.

The distinct difference between NP-$TiO_2$ and PN-$TiO_2$ suggests that the band gap reduction in NP-$TiO_2$ is a result of the defect levels introduced into the band gap by substitution of N and P at lattice sites, rather than other possible causes such as grain boundaries or disorders [13]. Otherwise, similar improvement in optical absorption should also be observed for PN-$TiO_2$. Because the VBM states are derived predominantly from O $2p$ states, N or P substitution for O mainly modifies the position of VBM. Thus, the band gap reduction in NP-TiO$_2$ is concluded to be a result of up-shift of VBM, which is a unique advantage of the all-anion codoping approach. The only possibility to introduce states near the CBM is P substitution for Ti. Our first-principles calculation, however, suggests that P$_{Ti}$ is a shallow donor and does not introduce any deep electronic levels.

Undisputable gap reduction by N-P chemical codoping is further evidenced by STS. Most of the measured current-voltage (I-V) curves are non-linear, with negligible current at low bias voltages, reflecting the semiconducting nature of the samples. Figures 4(a)-4(d) show the typical I-V curves and differential tunneling conductance ($d$I/$d$V) spectra, respectively. The latter reflects the local density of states (LDOS) of samples. Both I-V and $d$I/$d$V curves for NP-TiO$_2$ are visibly narrower than those for undoped TiO$_2$, suggesting significant gap reduction. Quantitative band gaps of the samples were obtained by cutting the $d$I/$d$V curves with a threshold LDOS (2.5% of the peak value) and the difference in the positive and negative bias voltages was taken as the gap value. The threshold was so chosen that the most probable gap value is 3.0 eV for undoped TiO$_2$, consistent with reported values. Figure 4(e) and 4(f) shows the band gap mapping for nanoparticle films over an area of 1×1μm$^2$, for both pure and NP-TiO$_2$; the gap values are represented by color scales. From the histograms of the gap values shown in Figure 4(g) and 4(h), it can be seen that the peak gap values have been reduced from ~3.0 eV for undoped TiO$_2$ to 1.8 eV for NP-TiO$_2$, a 40% reduction. This suggests that N-P chemical codoping creates new and accessible electronic states in the energy gap of TiO$_2$, which further confirms UV-vis absorption results and our theoretical predictions. The distribution of gap values for NP-TiO$_2$, with a full-width-at-half-maximum of 1.5 eV is significantly broader than 0.7 eV for undoped TiO$_2$. This may originate from inhomogeneous doping concentration.

In summary, we have developed a sequential doping strategy to realize chemical codoping on anion sites in a semiconductor oxide. The formation of new chemical bonds between anion dopant pair in $TiO_2$ results in band gap reduction by ~40%. Various characterizations support theoretical finding that chemical codoping simultaneously controls the band gap and band alignment, which could be an important step forward in improving the general performance of electronic and optoelectronic materials and devices. Formation of covalent bonding between dopants in a solid state material is a novel phenomenon. Its impact goes beyond the band gap engineering of semiconductors and provides a new direction for tuning the electronic structure of solids.

**Acknowledgements**

This work was supported by US NSF DMR1104994, DMR1006286, NSF of China 51101095, NSFC for Distinguished Young Scholars 51025101 and "One Hundred Talented People" of Shanxi Province. The supercomputer time is provided by NERSC under the US DOE Contract No. DE-AC02-05CH11231.

# References

[1] H. Amano, M. Kito, K. Hiramatsu, and I. Akasaki, Jpn. J. Appl. Phys. **28**, L2112 (1989).

[2] S. Nakamura, M. Senoh, N. Iwasa, and S. Nagahama, Jpn. J. Appl. Phys. **34**, L797 (1995).

[3] D. D. Awschalom, R. Epstein, and R. Hanson, Sci. Am. **297**, 84 (2007).

[4] P. Wang, Z. R. Liu, F. Lin, G. Zhou, J. Wu, W. H. Duan, B. L. Gu, and S. B. Zhang, Phys. Rev. B **82**, 193103 (2010).

[5] R. Long and N. J. English, Chem. Phys. Chem. **12,** 2604 (2011).

[6] C. D. Valentin, G. Pacchioni, and A. Selloni, Chem. Mater. **17**, 6656 (2005).

[7] W. Y. Choi, A. Termin, and R. M. Hoffmann, J. Phys. Chem. **98**, 13669 (1994).

[8] A. Fuerte, M. D. Hernández-Alonso, A. J. Maira, A. Martínez-Arias, M. Fernández-García, J. C.Conesa, and J. Soria, Chem. Commun. **24**, 2718 (2001).

[9] R. Asahi, T. Morikawa, T. Ohwaki, K. Aoki, and Y. Taga, Science **293**, 269 (2001).

[10] S. U. M. Khan, M. Al-Shahry, and W. B. Ingler Jr., Science **297**, 2243 (2002).

[11] Y. Q. Gai, J. B. Li, S. S. Li, J. B. Xia, and S. H. Wei, Phys. Rev. Lett. **102**, 036402 (2009).

[12] W. G. Zhu, X. F. Qiu, V. Iancu, X. Q. Chen, H. Pan, W. Wang, N. M. Dimitrijevic, T. Rajh, H. M. Meyer, III., M. P. Paranthaman, G. M. Stocks, H. H. Weitering, B. H. Gu, G. Eres, and Z. Y. Zhang, Phys. Rev. Lett. **103**, 226401 (2009).

[13] X. B. Chen, L. Liu, P. Y. Yu, and S. S. Mao, Science **331**, 746 (2011).

[14] A. Kudo, Catal. Surv. Asia **7**, 31 (2003).

[15] A. Hameed and M. A. Gondal, J. Mol. Catal. A **219**, 109 (2004).

[16] A. Fujishima and K. Honda, Nature **238**, 37 (1972).

[17] M. R. Hoffmann, S. T. Martin, W. Y. Choi, and D. W. Bahnmann, Chem. Rev. **95**, 69 (1995).

[18] S. Sakthivel and H. Kisch, Chem. Phys. Chem. **4**, 487 (2003).

[19] H. Q. Jiang, P. P. Yan, Q. F. Wang, S. Y. Zang, J. S. Li, and Q. Y. Wang, Chem. Eng. J. **215-216**, 348 (2013).

[20] B. L. Bischoff and M. A. Anderson, Chem. Mater. **7**, 1772 (1995).

[21] J. Yuan, M. X. Chen, J. W. Shi, and W. F. Shangguan, Int. J. Hydrogen Energ. **31**, 1326 (2006).

[22] A. E. Henkes and R. E. Schaak, Chem. Mater. **19**, 4234 (2007).

[23] J. Heyd, G. E. Scuseria, and M. Ernzerhof, J. Chem. Phys. **118**, 8207- (2003).

[24] G. Kresse and J. Furthmüller, Comput. Mater. Sci. **6**, 15 (1996).

[25] L. Körösi, S. Papp, I. Bertóti, and I. Dékány, Chem. Mater. **19**, 4811 (2007).

[26] R. Y. Zheng, L. Lin, J. L. Xie, Y. X. Zhu, and Y. C. Xie, J. Phys. Chem. C **112**, 15502 (2008).

[27] L. Lin, R. Y. Zheng, J. L. Xie, Y. X. Zhu, and Y. C. Xie, Appl. Catal. B-Enviro. **76**, 196 (2007).

[28] G. S. Shao, F. Y. Wang, T. Z. Ren, Y. P. Liu, and Z. Y. Yuan, Appl. Catal. B-Enviro. **92**, 61 (2009).

[29] J. Ananpattarachai, P. Kajitvichyanukul, and S. Seraphin, J. Hazard. Mater. **168**, 253 (2009).

[30] J. A. Rodriguez, T. Jirsak, J. Dvorak, S. Sambasivan, and D. Fischer, J. Phys. Chem. B **104**, 319 (2000).

[31] X. B. Chen and C. Burda, J. Phys. Chem. B **108**, 15446 (2004).

[32] Handbook of the elements and native oxides, XPS international, Inc. (1999).

[33] S. J. Splinter, R. Rofagha, N.S. Mcintyre, and U. Erb, Surf. Interface Anal. **24**, 81 (1996).

[34] M. Sathish, B. Viswanathan, R. P. Viswanath, and C. S. Gopinath, Chem. Mater. **17**, 6349 (2005).

**Figure captions:**

FIG. 1 (color online). (a) Atomic structure of an N-P pair substituting for two O atoms. (b) A schematic showing the bonding mechanism of the N-P pair. The lower (upper) gray areas represent the valence (conduction) band of undoped $TiO_2$. (c) and (d) show the charge density isosurfaces of the lower and higher gap states, respectively, of the N-P pair. Both orbitals show an anti-bonding character between P *3p* and N *2p* orbitals. To clearly show the *p-p* anti-bonding, the orientations of atomic structures in (c) and (d) are different from that in (a). An N-P-Ti triangle in pink color has been used to guide the eye.

FIG. 2 (color online). XPS of core level electron peaks of N *1s* and P *2p*. (a) N *1s* peak of NP-$TiO_2$, N-$TiO_2$, and PN-$TiO_2$. (b) P *2p* peak of NP-$TiO_2$, P-$TiO_2$, and PN-$TiO_2$. The dots are raw XPS data and the lines are fittings, respectively.

FIG. 3 (color online). (a) Experimental UV-vis absorption spectra of $TiO_2$, N-$TO_2$, P-$TiO_2$ and NP-$TiO_2$. (b) Calculated imaginary part ($\varepsilon_2$) of the dielectric constant for undoped $TiO_2$, N-P chemical codoped $TiO_2$, where both N and P substitute for O, and (N, P) compensated codoped $TiO_2$, where N substitutes for O and P substitutes for Ti. The dopant concentration is one N-P pair per 96-atom supercell (i.e., about 1% N and 1% P).

FIG. 4 (color online). Typical I-V curves and *dI/dV* spectra of undoped $TiO_2$ (a, c) and NP-$TiO_2$ (b, d). From representative regions, spatial mapping of the band gaps in an area of 1 um×1 um in size (gap values are represented by color scales) and the corresponding histograms of the band gaps for undoped $TiO_2$ (e, g) and NP-$TiO_2$ (f, h).

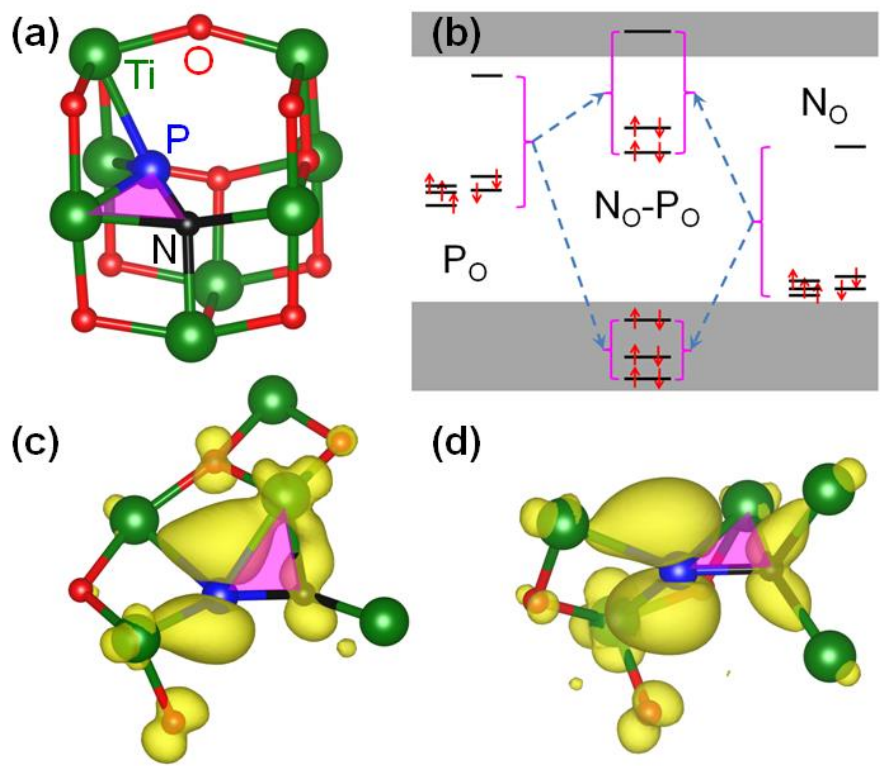

FIG. 1

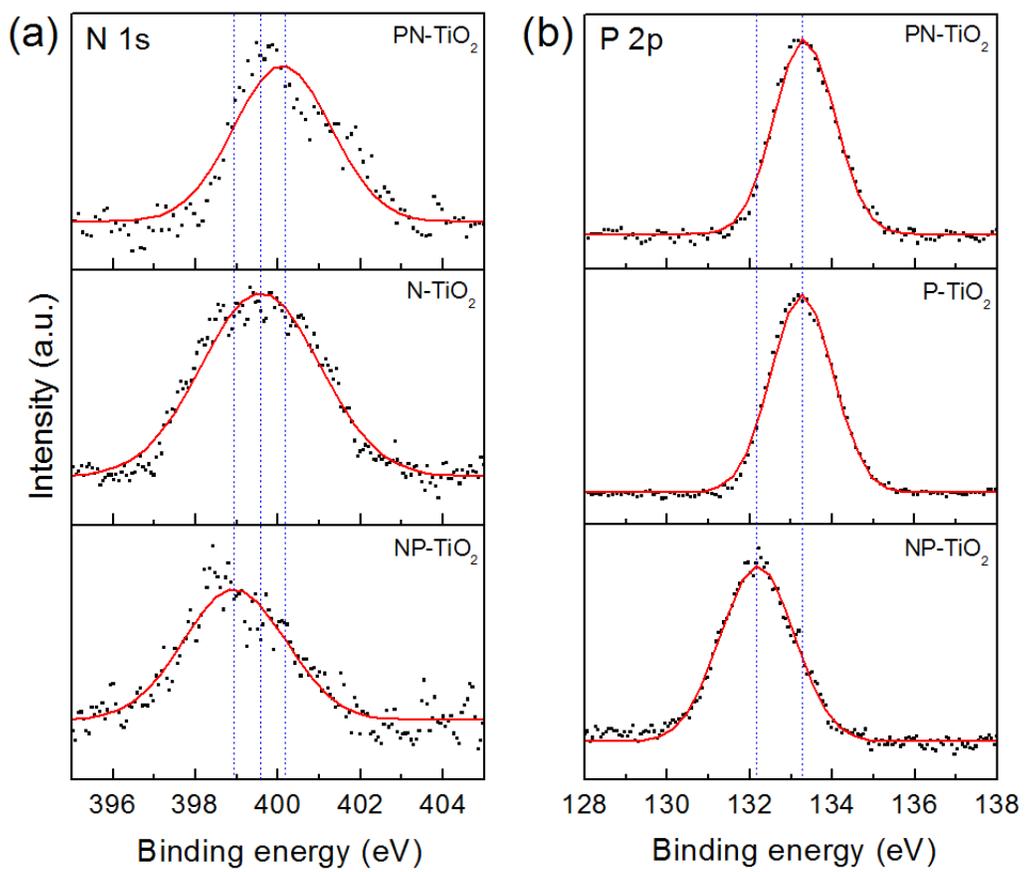

FIG. 2

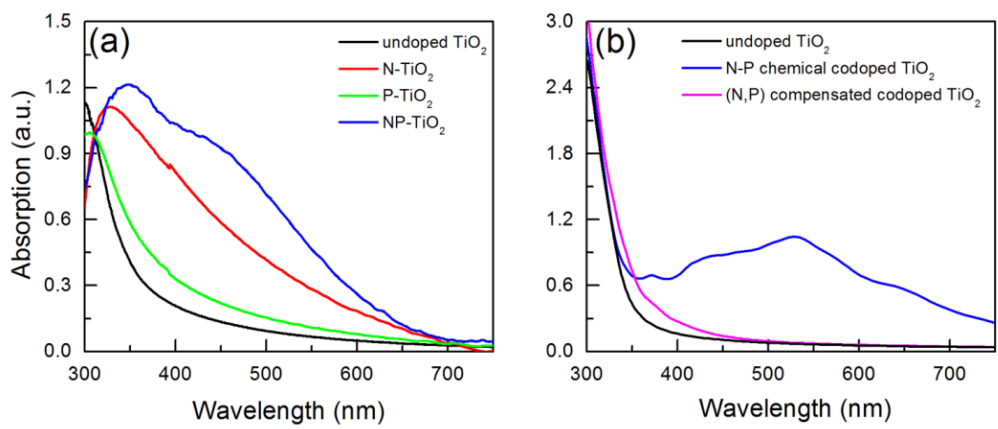

FIG. 3

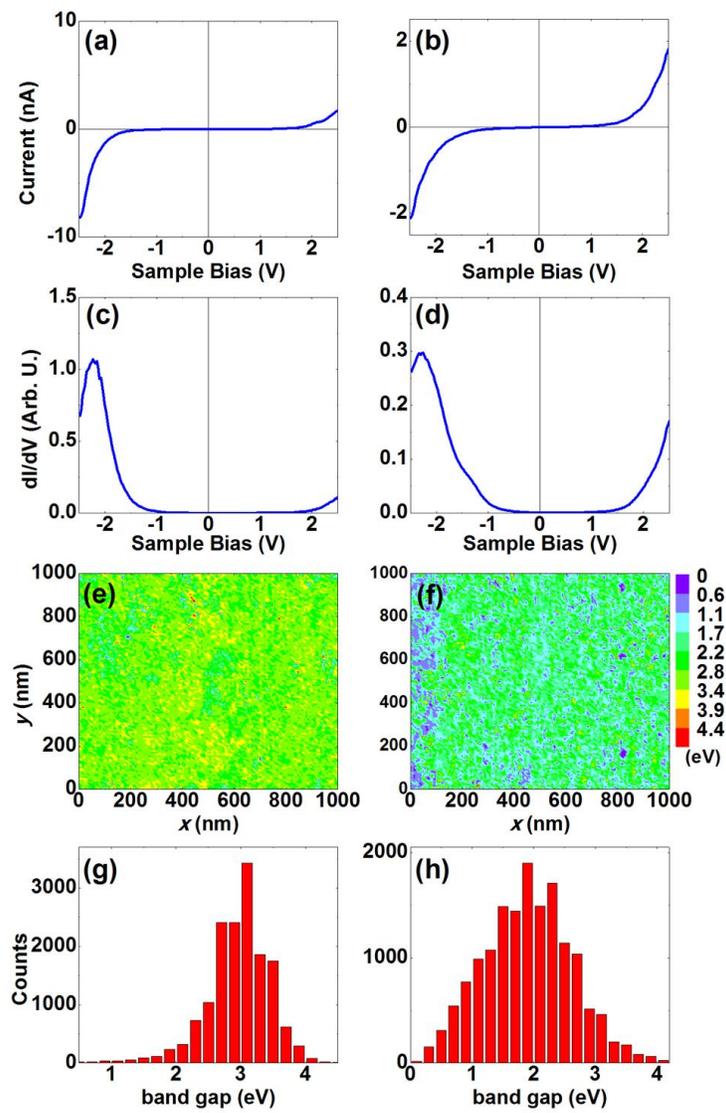

FIG. 4